**Treat-through OLED Displays: Dosimetry and performance of OLED AVATAR screens for Megavoltage Radiotherapy.**

Daniyal S Khan[1], Aaron Garza[1], Joseph B Schulz[1,2], Billy W Loo[1], Susan M Hiniker[1] Clinton Gibson[1], Lawrie B Skinner[1]

[1]Department of Radiation Oncology, Stanford University School of Medicine, 875 Blake Wilbur Drive, Stanford, CA 94305, United States

[2]Department of Human Radiation Oncology, University of Wisconsin Madison, 600 Highland Avenue, Madison, WI 53792 United States

**Author Contributions:**
Daniyal S Khan – Data collection, data analysis, manuscript drafting and review.
Aaron Garza – Device development and testing, manuscript review.
Joseph B Schulz – Device testing, manuscript review.
Billy W Loo – Concept input, clinical feedback, manuscript review.
Susan M Hiniker - clinical feedback, manuscript review.
Clinton Gibson – Device testing, manuscript review.
Lawrie B Skinner – Concept, initial device development and testing, data collection, manuscript drafting and review.

**Conflict of interest:** Lawrie B Skinner, Susan M Hiniker and Billy W Loo are listed as co-inventors of a Stanford University licensable technology packet related to AVATAR systems for radiotherapy.

**Acknowledgements:** Dr Sarah Susan Donaldson, provided feedback and discussion that lead to the development of this device.

**Treat-through OLED Displays: Dosimetry and performance of OLED AVATAR screens for Megavoltage Radiotherapy.**

**Objective**
The AVATAR (Audio Video Assisted Therapeutic Ambiance for Radiotherapy) system utilizes a radiolucent video display to help relax and immobilize pediatric patients during radiotherapy. This study investigates the use of OLED (Organic Light Emitting Diode) displays, which offer superior image quality and faster alignment compared to traditional projector-based systems, albeit with slightly increased thickness. The dose perturbations caused by these screens were assessed to evaluate their suitability for radiotherapy.

**Materials and Methods**
An 8" (20 cm) OLED screen was positioned in the patient's line of sight, between the patient's head and the radiation source. The screen comprises a 0.25 mm thick flexible OLED panel and a custom carbon fiber backing (0.3 to 0.6 mm thick), with most electronic components relocated via ribbon cable. 6 MV portal images were captured with and without the screen to create transmission maps. Additionally, parallel plate ion chamber measurements were taken at the surface and at depth in a solid water phantom. The OLED's radiation tolerance was tested with a 100 Gy dose delivery and 12 months of use in an active radiotherapy linac vault.

**Results**
The portal image transmission maps indicated an average attenuation difference of 0.3% (SD = 0.37%) between measurements with and without the AVATAR screen, with a maximum point attenuation of 3%. Ion chamber measurements revealed a 0.31% decrease in dose at 2 cm depth and a 7.04% increase at the surface under an anterior-posterior beam. For a VMAT arc, the dose difference was 0.04% at 2 cm depth and 0.15% at the phantom surface.

**Conclusions**
The minimal dose deviations indicate that these screens are suitable for megavoltage arc therapy without significantly affecting the target dose. While surface dose is slightly increased for static anterior beams, it remains low. For lower photon energies or particle beams, larger perturbations may occur, necessitating further measurements or beam entry avoidance.

# 1. Introduction

While radiotherapy is an effective cancer treatment, communication of information to patients inside the radiation shielding vault is often limited. Additionally, for children, radiotherapy can be stressful as they must remain still. This work builds on a system developed at Stanford university called AVATAR (Audio Video Assisted Therapeutic Ambiance for Radiotherapy) to display audio-visual content to pediatric patients during radiotherapy to help them stay calm and still [1-6]. These systems employ a projector with a radiotransparent screen. The projector-based setup, while effective, requires time to align and focus, has relatively low resolution, and the image is often distorted.

Paper-thin flexible Organic Light Emitting Diode (OLED) displays can potentially offer similar radio transparency while achieving improved image quality, faster setup time, and more setup flexibility. Radiation damage, however, is a concern for the use of OLED displays in a radiation environment. Here, the robustness and radio transparency of OLED screens are investigated for for megavoltage X-ray radiotherapy applications.

## 2. Materials and Methods

### 2.1 AVATAR system design and setup

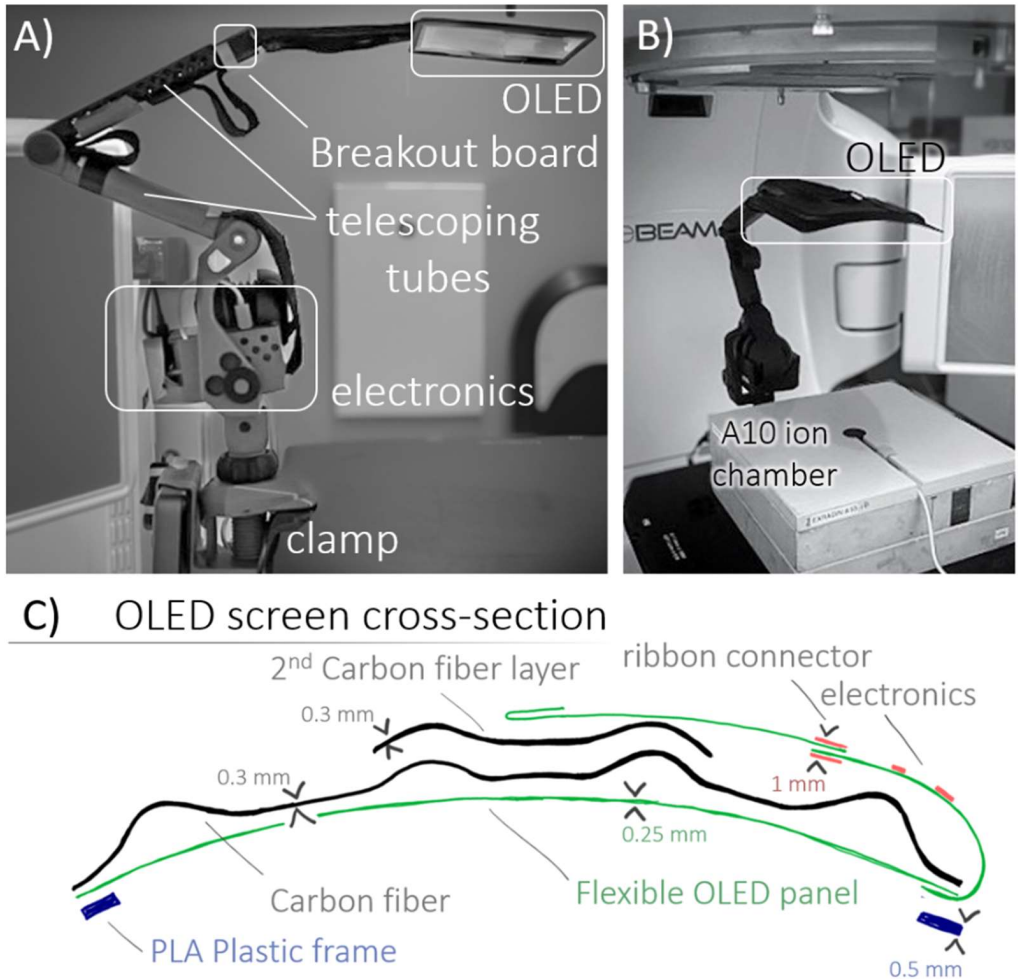

**Fig. 1.** A) AVATAR system setup for Electronic Portal Imaging device (EPID) measurement. B) Setup for the parallel plate ion chamber measurement, with the solidwater phantom and a source-surface distance of 100cm and the AVATAR screen at a distance of 70 cm from the radiation source (i.e. 30cm above the phantom). C) Cross-section schematic of the components within the OLED screen that are within the radiation field. The parts inside the radiation field include: carbon fiber reinforced plastic, a 3D printed polylactic acid frame, the flexible OLED panel and the ribbon data and power cable (see also Figure 2B). The experimental OLED setup demonstrates the AVATAR screen positioned in the patient's line of sight, 30 cm above the measurement phantom, allowing assessment of dose perturbations during radiotherapy.

The OLED AVATAR system was set up by clamping it to the side of the radiotherapy treatment couch and then adjusting the arms so that the screen was above the patient's face at a distance between 10 cm and 40 cm. The OLED AVATAR screen consists of a 0.25mm thick flexible OLED panel that is glued to a custom molded carbon fiber reinforced plastic backing. The carbon-fiber piece was 0.3 mm thick in most areas apart from the center, inferior region, which was 0.6mm for added strength. Taking a density of 1.5 g/cc for the carbon fiber parts and 1.3 g/cc for the OLED panel, yields an expected water equivalent thickness (WET) of 0.8-1.3 mm for the bulk of the screen area. The carbon fiber was molded to have a 60cm diameter curvature to minimize collision risk with the gantry. Grooves were added to increase stiffness without adding extra thickness. The carbon-fiber part is then connected to the 3D-printed telescoping arm. To minimize electronics in the beam, the main high-definition multimedia interface (HDMI) driver board and speaker components are placed at the base of the 3D printed mount and connected via ribbon cables. There remain some small electronic components and the ribbon cable connector over the screen on a flexible tape. The thickest part of the screen in the beam is the ribbon cable connector which is two stacked 15x5x0.5 $mm^3$ steel plates with a mix of plastic and copper in between (Fig. 1C, and Fig. 2). Finally, there is a small breakout electronics circuit board near the top of the telescoping arms 20cm inferiorly down from the center of the screen, which should remain outside of direct radiation beams.

## 2.2 Portal image transmission maps

The AVATAR system was set up as shown in Fig 1A, with the screen placed in the beam, 30 cm above the linac's isocenter. During portal imaging, there was nothing placed on the treatment couch. A 50 MU (Monitor Unit) 6 MV X-ray beam was delivered through the screen and couch, which was imaged with an electronic portal imager (EPID, model AS1200) to obtain an integrated image with intensities I(x,y). A second otherwise identical image was then recorded without AVATAR in the beam to obtain $I_0(x,y)$. The attenuation of the AVATAR system was then obtained from a ratio image, RI(x,y) from the ratio of the I and $I_0$ EPID images using ImageJ software. This RI image is a 2D projection of the transmission of the 6MV beam through the AVATAR screen (eqn. 1).

$$RI(x,y) = I(x,y) / I_0(x,y) \quad \text{eqn. 1}$$

## 2.3 Ion Chamber measurements

The AVATAR system was set up for ion chamber measurements as shown in Fig 1B. A 30 x 30 x 12 $cm^3$ solid water phantom, with an Exradin A10 parallel plate ion chamber (collector diameter, 5.4 mm, plate separation 2.0 mm, 0.0386 mm of water equivalent buildup) was placed underneath the OLED screen at a Source to phantom surface distance of 100 cm. The ion chamber was placed at depths of 0, 2, and 5 cm in the phantom. Three beam arrangements were tested: (i) A simple 20x20 cm2 open field anterior-posterior beam. (ii) An anterior half-arc, at 6MV with 200MU at 600MU/min from gantry angles 270 to 90 degrees, with a conformal Multi-Leaf Collimator (MLC) pattern tracking a 10x10 cm target. (iii) A full 360-degree 6MV VMAT arc taken from an anonymized Glioblastoma multiforme (GBM) treatment plan, with 351 MU spread over the full 360-degree arcs.

For each beam type, the deliveries were repeated with and without the AVATAR screen, and the dose difference ΔD, was calculated from the difference of the dose with the screen, $D_S$ and the dose without the screen, $D_N$ normalized by $D_N$, (equation 2).

$$\Delta D = |D_S - D_N| / D_N \quad \text{eqn. 2.}$$

### 2.4 Radiation tolerance testing

For radiation tolerance testing, a 20x20 cm$^2$ wide 6 MV radiation field was delivered to the screen at a source to surface distance of approximately 70cm. The dose rate at the water equivalent depth of 0.5-1.0 mm is 40-50% relative to the depth of maximum dose. The 70cm distance means an inverse square factor of 2 compared to the 100 cm isocenter distance. The dose from the 10,000 MU delivered to the OLED pane, was therefore in the range 80-100 Gy.

## 3. Results

### 3.1 Portal image transmission results

Analysis of the screen region and background regions of the transmission map from 6 MV portal imaging is shown in Figure 2. The OLED screen region had a transmission value of 99.6% with a standard deviation of 0.37%. The 1 sigma statistical uncertainty in these measurements was found to be +-0.14%. This uncertainty was obtained from the standard deviation in the background region of the ratio image, which contains only measurement noise. For completeness, the mean transmission in the background region was 100.1%. There was a small area, approximately 15x5 mm, with a maximum attenuation of 3%. The areas of highest attenuation were the ribbon cable connector (indicated in Figure 2) and the region where the flexible electronic connections were folded over. The ribbon cable has two 15 x 5 x 0.4 mm steel plates, and a 1mm thick plastic connector in between. Including the OLED and carbon fiber layers and using a relative electron density of 6.5 for steel, gives an approximate water equivalent thickness of ~8mm at the location of the ribbon cable connection, compared to 0.8-1.2 mm WET in the majority of the screen.

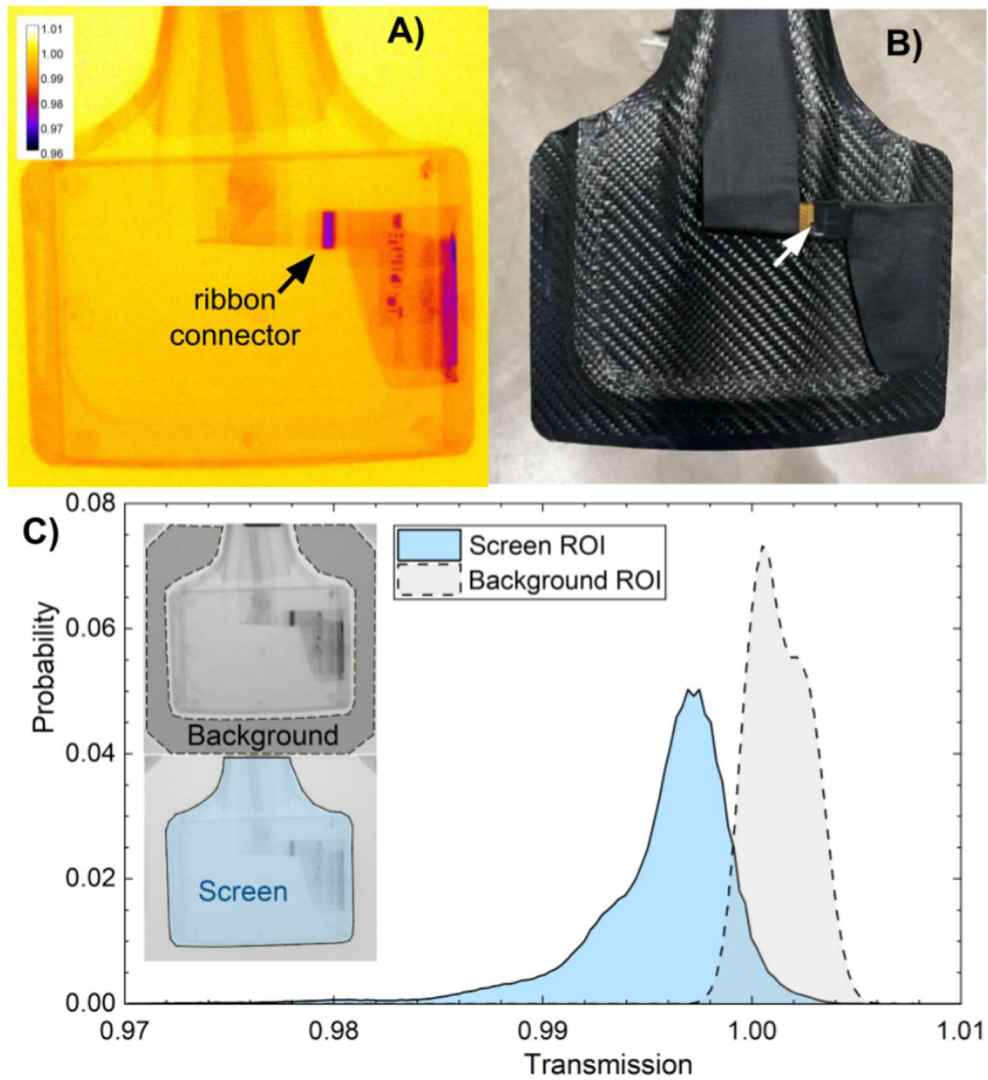

**Fig. 2:** A) Ratio image showing the measured transmission map of the AVATAR device in a 6 MV beam. The image shows that the fraction of the beam transmitted through the screen region of interest (ROI) is in the range 0.96-1.0, with average transmission 0.996. i.e., 0.4% mean attenuation. B) Image of the OLED screen. C) The attenuation of the screen, coupled with the error of said measurements. The OLED screen shows minimal beam attenuation, with 99.6% average transmission across most of the screen area.

|  | Mean (transmission) | Standard Deviation |
|---|---|---|
| Screen ROI | 99.6% | 0.37% |
| Background ROI | 100.1% | 0.14% |

Table. 1: Analysis of the two regions of interest (ROI) in the Ratio Image (transmission map) obtained from Portal imaging.

## 3.2 Ion chamber results

|  | Surface Dose difference (%) | 2-cm-depth dose difference (%) | 5-cm-depth dose difference (%) |
|---|---|---|---|
| **Single AP beam** | 7.04% | -0.31% | -0.35% |
| **Half arc** (90 to 270) (3D conformal) | 1.60% | -0.01% | - |
| **Full VMAT arc** | 0.15% | 0.04% | - |

Table. 2: Dose differences measured with an Exradin A10 parallel plate ion chamber due to the presence of the OLED AVATAR screen for 6 MV beams. Values represent percentage differences between dosage and without the AVATAR system, calculated from raw electrometer data (2 trials per setup). Measurement uncertainty was approximately 0.1%. Surface dose increases significantly for single anterior-posterior (AP) beams (7.04%) but decreases to levels <2% for arc therapy, while dose at depth remains within 0.35% for all beam configurations. Measured differences between radiation deliveries with and without the AVATAR screen in the beam. uncertainty in the measurements was approximately 0.1%

As shown in Table 2, for a single AP beam, the surface radiation dose increased by about 7%, whereas at 2 cm and 5 cm deep, the radiation dose decreased by about 0.3%. For arc therapy beams however, the dose deviations were all less than 2% at all measured depths.

# 4. Discussion

## 4.1 Impact of radiation on the OLED screen

Radiation tolerance testing by delivering 10,000 MU (approximately 100 Gy) of 6MV radiation through the screen resulted in no signal breakup and no noticeable dimming of the OLED screen. This is consistent with literature for space applications, where it was reported that approximately 1% brightness loss occurs per 200 Gy of radiation [7].

Further, clinical use of 2 devices in radiotherapy vaults over a 1-year period saw minimal radiation-induced degradation of the screens. Note that given the small angular coverage of the 20 cm screen, each volumetric modulated arc therapy (VMAT) treatment arc will result in approximately 50 MU (~0.5 Gy) or less being sent through the screen per arc. This would be an expected dimming of 0.0025% per treatment arc. Of note was image burn-in, which was significant where screens were used with the Varian visual feedback system over several months (as this is a largely static image without any screen-saving features). In clinical use, the most common issues found were with the ribbon cable connections during and after mechanical adjustment of the telescoping screen arm. This was addressed with a combination of stoppers on the telescope adjustment, as well as hot glue, tape, and plastic covers over connections. Intermittent electrical malfunctions requiring system power reset occurred approximately 1-2 times per month; however, the contribution of radiation exposure to these malfunctions could not be determined.

## 4.2 Impact of the OLED screen on radiotherapy dose

The 6 MV portal images and the ion chamber dose attenuation measurements demonstrated an average 6MV beam attenuation of 0.3 to 0.4% from the OLED screens (Table 3). With a small area receiving approximately up to 3% attenuation (Figure 2, Tables 1, 3). For VMAT arcs, where the dose is more spread out, this was deemed clinically acceptable. Skin dose is more significantly increased for a single beam, whereas for VMAT arcs, the resulting dose increase on the patient surface was clinically insignificant. Given the low dose perturbation and variable screen position, it is generally not helpful to include the screen in the dose calculation. Care should be taken, however, for static photon beam treatments through the screen, as the added dose to the lens of the eye may be a concern in some situations. NThis 7% additional surface dose is far less than the amount expected from a typical thermoplastic mask over the eye, where the dose increase from the presence of the mask is in the range 200-400% [8].

# 5. Conclusion

The durability and radio transparency demonstrated here indicates that OLED screens such as these minimally impact radiation dose, while the radiation damage has minimal impact on the display. These OLED screens, therefore, are potentially a useful and reliable communication and relaxation tool for megavoltage radiotherapy linac treatments.

# References

1. Schulz JB, Zalavari L, Gutkin P, Jiang A, Wang YP, Gibson C, Garza A, Bush KK, Wang L, Donaldson SS, Loo BW, Hiniker SM, Skinner L. AVATAR 2.0: next level communication systems for radiotherapy through face-to-face video, biofeedback, translation, and audiovisual immersion. Front Oncol. 2024 Oct 8;14:1405433. doi: 10.3389/fonc.2024.1405433.
https://doi.org/10.3389/fonc.2024.1405433

2. Paulina M. Gutkin, Lawrie Skinner, Alice Jiang, Sarah S. Donaldson, Billy W. Loo, Justin Oh, Yi Peng Wang, Feasibility of the Audio-Visual Assisted Therapeutic Ambience in Radiotherapy (AVATAR) System for Anesthesia Avoidance in Pediatric Patients: A Multicenter Trial, International Journal of Radiation Oncology*Biology*Physics, Volume 117, Issue 1, 2023, Pages 96-104, ISSN 0360-3016, - https://doi.org/10.1016/j.ijrobp.2023.03.063.

3. Katy E. Balazy, Paulina M. Gutkin, Lawrie Skinner, Rie von Eyben, Tyler Fowler, Daniel W. Pinkham, Susan M. Hiniker, Impact of Audiovisual-Assisted Therapeutic Ambience in Radiation Therapy (AVATAR) on Anesthesia Use, Payer Charges, and Treatment Time in Pediatric Patients, Practical Radiation Oncology, Volume 10, Issue 4, 2020, Pages e272-e279, ISSN 1879-8500, https://doi.org/10.1016/j.prro.2019.12.009.

4. Susan M. Hiniker, Karl Bush, Tyler Fowler, Evan C. White, Samuel Rodriguez, Peter G. Maxim, Initial clinical outcomes of audiovisual-assisted therapeutic ambience in radiation therapy (AVATAR), Practical Radiation Oncology, Volume 7, Issue 5, 2017, Pages 311-318, ISSN 1879-8500, https://doi.org/10.1016/j.prro.2017.01.007.

5. Rahul N. Prasad, Sujith Baliga, Julie Banner, Catherine Cadieux, Ashley Cetnar, Michael Degnan, Radiation Therapy Without Anesthesia for a 2-Year-Old Child Using Audio-Visual Assisted Therapeutic Ambience in Radiation Therapy (AVATAR), Practical Radiation Oncology, Volume 12, Issue 3, 2022, Pages e216-e220, ISSN 1879-8500, https://doi.org/10.1016/j.prro.2021.12.009.

6. Shearwood McClelland, Kent W. Overton, Brian Overshiner, Karl Bush, Billy W. Loo, Lawrie B. Skinner, Gordon A. Watson, Jordan A. Holmes, Susan M. Hiniker, Peter G. Maxim, Cost Analysis of Audiovisual-Assisted Therapeutic Ambiance in Radiation Therapy (AVATAR)-Aided Omission of Anesthesia in Radiation for Pediatric Malignancies, Practical Radiation Oncology, Volume 10, Issue 2, 2020, Pages e91-e94, ISSN 1879-8500, doi: 10.1016/j.prro.2019.09.011.

7. L. D. Ryder *et al*., "An Examination of the Radiation Sensitivity of Electronic Display Pixel Technologies," *2023 IEEE Radiation Effects Data Workshop (REDW) (in conjunction with*

*2023 NSREC)*, Kansas City, MO, USA, 2023, pp. 1-7, doi: 10.1109/REDW61050.2023.10265850. https://www.doi.org/10.1109/REDW61050.2023.10265850

8. Pourfallah T, Pirzadeh M, Seifi Makrani D, Mihandoust E, Davodian S. Surface and the Build-up Dose Distribution Due to Applying Thermoplastic Masks in External Radiotherapy. Frontiers Biomed Technol. 2020;7(4):217-225. doi: 10.18502/fbt.v7i4.5319 [https://doi.org/10.18502/fbt.v7i4.5319]

## Supplemental material

|  | No Avatar screen (pC) | AVATAR screen (pC) |
|---|---|---|
| Surface | 341, 341 | 365, 365 |
| 2 cm deep | 1641, 1641 | 1636, 1636 |
| 5 cm deep | 1431, 1432 | 1426, 1427 |

**Table. 3:** Raw parallel plate ion chamber data for single AP beam, collected from the electrometer (2 trials for each setup).

|  | No Avatar screen (pC) | AVATAR screen (pC) |
|---|---|---|
| Surface | 1496.0 | 1520.0 |
| 2 cm deep | 3294.4 | 3294.1 |

**Table. 4:** Raw parallel plate ion chamber data for half arc (90 to 270 degrees), collected from the electrometer.

|  | No Avatar screen (pC) | AVATAR screen (pC) |
|---|---|---|
| Surface | 610.71 | 611.61 |
| 2 cm deep | 1473.6 | 1474.2 |

**Table. 5:** Raw parallel plate ion chamber data for a full VMAT arc, collected from the electrometer.